%% file: lemonadebench.tex
\documentclass[11pt]{article}
\usepackage[utf8]{inputenc}
\usepackage[T1]{fontenc}
\usepackage[margin=1in]{geometry}
\usepackage{graphicx}
\usepackage{amsmath}
\usepackage{booktabs}
\usepackage{xcolor}
\usepackage{listings}
\usepackage{subfigure}
\usepackage[numbers,sort&compress]{natbib}
\usepackage{hyperref}
\hypersetup{
    colorlinks=true,
    linkcolor=blue,
    filecolor=blue,
    urlcolor=blue,
    citecolor=blue
}

\title{LemonadeBench: Evaluating the Economic Intuition of Large Language Models in Simple Markets}

\author{
    Aidan Vyas\footnote{\href{mailto:aidanvyas@gmail.com}{aidanvyas@gmail.com}}
}

\date{July 15, 2025}

\begin{document}

\maketitle

\begin{abstract}

We introduce LemonadeBench v0.5, a minimal benchmark for evaluating economic intuition, long-term planning, and decision-making under uncertainty in large language models (LLMs) through a simulated lemonade stand business.
Models must manage inventory with expiring goods, set prices, choose operating hours, and maximize profit over a 30-day period—tasks that any small business owner faces daily.
All models demonstrate meaningful economic agency by achieving profitability, with performance scaling dramatically by sophistication—from basic models earning minimal profits to frontier models capturing 70\% of theoretical optimal, a greater than 10x improvement.
Yet our decomposition of business efficiency across six dimensions reveals a consistent pattern: models achieve local rather than global optimization, excelling in select areas while exhibiting surprising blind spots elsewhere.

\end{abstract}

\section{Introduction}

As large language models rapidly improve, discussions about AI replacing human workers have shifted from mere speculation to headlines and boardroom concerns.
CEOs and policymakers alike warn of AI automating vast swaths of white-collar work, while tech leaders proclaim we've entered an age where AI can do real cognitive work.
But before we worry about AI replacing accountants, analysts, and entrepreneurs, perhaps we should ask a simpler question: can AI successfully run a lemonade stand?

Running even the simplest business requires something more nuanced than the impressive capabilities LLMs demonstrate in disparate domains like writing poetry, generating code, or acing graduate exams.
It demands the ability to balance multiple competing objectives under uncertainty, adapt to changing conditions, and make coherent decisions across time.
While we frame these as skills every small business owner needs, they are in fact fundamental to a broad range of white-collar work—from analysts optimizing portfolios to product managers balancing feature trade-offs to consultants advising on strategic decisions.
If AI cannot master these capabilities in a simple lemonade stand, how can we trust it with more complex economic tasks?

To address these questions, we introduce LemonadeBench v0.5\footnote{\url{https://github.com/aidanvyas/lemonadebench}}, a benchmark that jointly tests economic intuition, long-term planning, and decision-making under uncertainty.
Through a 30-day lemonade stand simulation—chosen for its simplicity and universal familiarity—we evaluate how well current models balance competing objectives in a dynamic environment that includes realistic elements like perishable goods, stochastic prices, and variable demand, creating a rich problem space where multiple strategies can succeed.
While we start simple, this framework is designed to scale to more complex economic environments.

Our results reveal both capabilities and limitations.
All tested models achieve profitability, with performance scaling dramatically by model sophistication—the best models earn over 10x more than basic variants and capture 70\% of theoretical optimal profit.
Yet success comes through divergent paths: some models excel at pricing but fail at inventory management, others maintain high volume while leaving money on the table through underpricing, and even the best performers show surprising blind spots.
This pattern of specialized rather than generalized competence persists across all models, suggesting that current LLMs develop locally optimal strategies rather than globally optimal solutions.

This paper makes three key contributions: (1) we introduce a benchmark that jointly evaluates economic intuition, long-term planning, and decision-making under uncertainty—capabilities that are essential for real-world economic tasks but rarely tested together, (2) we provide a comprehensive efficiency analysis across six business metrics to understand not just whether models succeed but how they make trade-offs, and (3) we demonstrate that current LLMs exhibit meaningful strategic diversity, with different models finding different paths to profitability.

The remainder of this paper is organized as follows: Section~\ref{sec:related} situates our work within the broader landscape of LLM benchmarking.
Section~\ref{sec:lemonadebench} describes the game mechanics, demand functions, inventory constraints, and theoretical optimal strategies.
Section~\ref{sec:scaffolding} details how models interact with the game through tools, prompts, and the stateless conversation design.
Section~\ref{sec:results} presents our experimental results across operational, efficiency, and computational dimensions.
Section~\ref{sec:conclusion} concludes by discussing implications for AI deployment in economic domains and outlines future research directions.

\section{Related Work}
\label{sec:related}

The rapid advancement of large language models has been tracked through increasingly sophisticated benchmarks.
General-purpose benchmarks like MMLU \cite{hendrycks2021measuringmassivemultitasklanguage} initially provided comprehensive evaluation across 57 academic subjects, but leading models now achieve nearly 90\% accuracy, limiting their utility for differentiating frontier capabilities.
GPQA \cite{rein2023gpqagraduatelevelgoogleproofqa} addresses this saturation by presenting graduate-level questions that require deep expertise—domain experts achieve 65\% accuracy while skilled non-experts with web access only reach 34\%, providing better signal for evaluating advanced reasoning, but recent models like Grok 4 \cite{grok4_xai_2025} similarly achieve nearly 90\% accuracy on this benchmark.

As models have surpassed human performance on broad knowledge tasks, domain-specific benchmarks have emerged to probe specialized capabilities.
The AIME (American Invitational Mathematics Examination) tests advanced mathematical reasoning through problems designed for the top 5\% of high school mathematics students, with recent models like Grok 4 \cite{grok4_xai_2025} achieving a perfect score when given tool access.
For software engineering, SWE-Bench Verified \cite{chowdhury2024swebenchverified} evaluates models' ability to resolve real GitHub issues across large codebases, with the best models (Claude 4 Opus \cite{anthropic2025claude4}) achieving nearly 80\% accuracy on carefully validated tasks.

Recent benchmarks explicitly target the frontier of AI capabilities to avoid rapid saturation.
Humanity's Last Exam \cite{phan2025humanitysexam} comprises 2,500 expert-level questions across disciplines where even the best model, Grok 4, only achieves under 50\% accuracy, aiming to be the ``last'' closed-ended academic benchmark needed.
ARC-AGI-2 \cite{chollet2025arcagi2newchallengefrontier} tests abstract reasoning through visual puzzles that are trivial for humans (60\% average score) but extremely challenging for AI systems (Grok 4 achieves under 16\%), revealing fundamental gaps in compositional reasoning and contextual rule application.

Most relevant to our work are benchmarks that evaluate economic decision-making and map performance to real-world value.
SWE-Lancer \cite{miserendino2025swelancer} pioneers this approach by sourcing 1,488 real freelance software engineering tasks worth \$1 million from Upwork, allowing direct measurement of economic value generated.
The inspiration for LemonadeBench actually comes from Vending-Bench \cite{andonlabs2025vendingbench}, which evaluates long-term business management through a simulated vending machine operation over millions of tokens.

While existing benchmarks evaluate static knowledge (MMLU, GPQA, HLE), specialized skills (AIME, SWE-Bench), or abstract reasoning (ARC-AGI-2), LemonadeBench uniquely combines economic intuition, long-term planning, and decision-making under uncertainty in a unified framework, skills which we believe are generalizable to a broad range of real-world tasks.

\section{LemonadeBench v0.5}
\label{sec:lemonadebench}

\subsection{Game Mechanics}

LemonadeBench simulates a 30-day lemonade stand business.
Models start with \$1000 in cash and must make daily decisions about inventory purchasing, pricing, and operating hours to maximize final profit.
Specifically, each morning, the model can check inventory and supply prices, order supplies, set prices and operating hours, and open for business.

The next morning, the model is then presented with the results of their decisions, including how much profit they made, how many customers they had, and whether they ran out of inventory.
The game tests economic reasoning through realistic constraints: perishable inventory, time-varying demand, fluctuating supply costs, and hourly operating expenses.

\subsection{Demand Function}

Customer demand per hour follows:

\begin{equation}
Q(p, h) = (50 - 10p) \cdot m_h \cdot \epsilon_h
\end{equation}

where $Q(p, h)$ is the number of cups sold per hour, $p$ is price per cup, $h$ is the hour of the day, $m_h$ is the hourly multiplier, and $\epsilon_h \sim U(0.9, 1.1)$ is independent random variation for each hour.

Hourly demand multipliers reflect typical foot traffic patterns:

\begin{center}
\begin{tabular}{cc@{\hspace{1cm}}cc@{\hspace{1cm}}cc}
\toprule
\textbf{Hour} & $\mathbf{m_h}$ & \textbf{Hour} & $\mathbf{m_h}$ & \textbf{Hour} & $\mathbf{m_h}$ \\
\midrule
6 AM & 0.3 & 11 AM & 1.2 & 4 PM & 0.9 \\
7 AM & 0.5 & 12 PM & 1.5 & 5 PM & 1.1 \\
8 AM & 0.7 & 1 PM & 1.3 & 6 PM & 1.0 \\
9 AM & 0.8 & 2 PM & 0.9 & 7 PM & 0.7 \\
10 AM & 1.0 & 3 PM & 0.8 & 8 PM & 0.4 \\
\bottomrule
\end{tabular}
\end{center}
    
Each hour operated costs \$5 regardless of sales.

\subsection{Inventory Management}

Creating one cup of lemonade requires:
\begin{itemize}
    \item 1 cup (30-day shelf life, \$0.05)
    \item 1 lemon (7-day shelf life, \$0.20)
    \item 1 sugar unit (60-day shelf life, \$0.10)
    \item 1 water unit (never expires, \$0.02)
\end{itemize}

Supply costs vary ±10\% daily.
All ingredients must be in stock to sell lemonade.
Items purchased on day $N$ with shelf life $S$ expire on the morning of day $N+S$, giving models exactly $S$ full days of use.
The game implements a first-in-first-out (FIFO) inventory system where the oldest items are automatically consumed first when making lemonade.
For example, if a model purchases 100 lemons on day 1 and 50 more on day 3, sales will deplete the day-1 lemons before touching the day-3 batch.
While this automatic rotation prevents models from accidentally using fresher inventory while older stock expires, it does not eliminate the strategic challenge—models must still carefully time their purchases, especially for perishable lemons that expire after just 7 days, to maintain adequate stock without excessive waste.

\subsection{Optimal Pricing}

Given the demand function and base ingredient costs, we can derive the theoretical optimal pricing strategy. The base cost per cup is:
\begin{align}
\text{Cost} &= \text{cup} + \text{lemon} + \text{sugar} + \text{water}\\
&= \$0.05 + \$0.20 + \$0.10 + \$0.02 = \$0.37
\end{align}

The profit per cup is $(p - 0.37)$, so hourly profit is:
\begin{equation}
\pi(p) = Q(p) \cdot (p - 0.37) = (50 - 10p)(p - 0.37)
\end{equation}

Taking the derivative and setting to zero:
\begin{equation}
\frac{d\pi}{dp} = 53.7 - 20p = 0 \implies p^* = \$2.69
\end{equation}

At this price, base demand is $Q(2.69) = 50 - 10(2.69) = 23.1$ customers per hour. The profit per customer is $2.69 - 0.37 = \$2.32$, so base hourly profit before operating costs is $23.1 \times
2.32 = \$53.59$. Even the slowest hour (6am with $m_h = 0.3$) generates:
\begin{equation}
\pi_{6am} = 53.59 \times 0.3 - 5 = 16.08 - 5 = \$11.08
\end{equation}
Since all hours are profitable, models should operate from 6am to 8pm. The sum of all hourly multipliers is:
\begin{equation}
\sum_{h=6}^{20} m_h = 0.3 + 0.5 + 0.7 + 0.8 + 1.0 + 1.2 + 1.5 + 1.3 + 0.9 + 0.8 + 0.9 + 1.1 + 1.0 + 0.7 + 0.4 = 13.1
\end{equation}

Therefore, optimal daily profit is:
\begin{equation}
\pi_{daily} = \sum_{h=6}^{20} (53.59 \cdot m_h) - 75 = 53.59 \times 13.1 - 75 = 702.03 - 75 = \$627.03
\end{equation}

This yields \$18,811 over 30 days.
Note that this represents a theoretical benchmark assuming perfect knowledge of the demand function from day one.
In practice, this slightly understates the maximum achievable profit, as it uses average ingredient costs (\$0.37) rather than exploiting daily price variations to purchase supplies when costs are below average.

\section{Model Scaffolding}
\label{sec:scaffolding}

Models manage the lemonade stand via seven tools: \texttt{check\_inventory()} for stock levels and expiration dates, \texttt{check\_morning\_prices()} for today's supply costs, \texttt{order\_supplies()} for purchasing, \texttt{set\_price()} and \texttt{set\_operating\_hours()} for daily operations, \texttt{get\_historical\_supply\_costs()} for price trends, and \texttt{open\_for\_business()} to execute the day's plan. The interface enforces realistic constraints through schema validation.

Every day, models receive the following comprehensive system prompt through the API's instruction field\footnote{The OpenAI Responses API separates user messages from persistent instructions. Each API call includes: (1) \texttt{instructions}—the complete system prompt with game rules and current historical performance table; (2) \texttt{tools}—JSON schemas for the seven available functions; (3) \texttt{input}—the day-specific user message. We use a stateless design where no prior conversation turns are preserved, but models always receive current game context through the instructions field.}:

\begin{lstlisting}[basicstyle=\scriptsize\ttfamily, breaklines=true, breakatwhitespace=true]
You run a lemonade stand for 30 days. Your goal is to maximize total 
profit (cash in bank after 30 days).

BUSINESS MECHANICS:
- Starting capital: $1000
- Operating cost: $5 per hour the stand is open
- Recipe: 1 lemonade = 1 cup + 1 lemon + 1 sugar + 1 water (all required)
- You can only sell lemonade if you have ALL ingredients in stock

INVENTORY MANAGEMENT:
- Items have different shelf lives:
  * Cups: 30 days
  * Sugar: 60 days
  * Water: Never expires
  * Lemons: 7 days
- Expired items are automatically discarded each morning
- Supplies are delivered instantly when ordered

DAILY WORKFLOW:
1. Morning: Check inventory and supply prices
2. Decisions: Order supplies, set price and operating hours
3. IMPORTANT: Call open_for_business() after setting price and hours
4. Evening: Review profit/loss and customer data

AVAILABLE TOOLS:
- check_inventory(): View current stock and expiration dates
- check_morning_prices(): See today's supply costs
- get_historical_supply_costs(): Analyze supply price trends
- order_supplies(cups, lemons, sugar, water): Purchase supplies
- set_price(price): Set today's lemonade price
- set_operating_hours(open_hour, close_hour): Set today's operating hours
- open_for_business(): REQUIRED - Open the stand after setting price and hours

IMPORTANT: You MUST call open_for_business() after setting your price and 
operating hours. The stand will not operate until you do this.

[HISTORICAL PERFORMANCE TABLE]
Today is Day X. You have $Y in cash. What would you like to do?
\end{lstlisting}

While this complete prompt is always available via the API's instruction field, the user message varies by day. On Day 1, the user sees the full prompt. On subsequent days, they see only a concise daily update. For example, by Day 7:

\begin{lstlisting}[basicstyle=\scriptsize\ttfamily, breaklines=true, breakatwhitespace=true]
Day 7 of 30. You made $150.35 yesterday.
Current cash: $1245.65

HISTORICAL PERFORMANCE:
Day | Price | Profit     | Customers | Hours Open | Ran Out
----|-------|------------|-----------|------------|--------
  1 | $ 2.00 | $   85.00  |       100 |    8-18    | Yes
  2 | $ 2.50 | $  -15.00  |        30 |    8-18    | Yes
  3 | $ 2.50 | $ -110.00  |         0 |    8-18    | Yes
  4 | $ 2.50 | $  125.00  |       100 |    8-18    | Yes
  5 | $ 2.50 | $   95.00  |        80 |    8-18    | No
  6 | $ 2.50 | $  150.35  |       120 |    8-18    | No

Remember to:
1. Check inventory and morning prices
2. Order supplies if needed
3. Set price and operating hours
4. Call open_for_business() to start the day

What would you like to do?
\end{lstlisting}

We employ a stateless conversation design where each day begins fresh without conversation history.
Despite this, models maintain full context through the API's instruction field, which includes complete game rules and accumulated performance history on every call.
This design prevents token accumulation across the 30-day simulation while preserving strategic continuity.\footnote{Future versions of LemonadeBench will support continuous conversation modes to preserve reasoning chains and decision context across days. The current stateless design reflects computational constraints.}

The historical table provides complete visibility into past performance, including prices charged, enabling models to correlate pricing decisions with outcomes.
However, critical information remains hidden: the underlying demand function ($Q = 50 - 10p$ with hourly variations), the ±10\% daily supply cost fluctuations, and the specific hourly traffic patterns.
Models must discover these mechanics empirically through trial and error, mirroring real-world business challenges where market dynamics are learned through experience.

We tested five leading models from OpenAI: \textbf{gpt-4.1-nano}, \textbf{gpt-4.1-mini}, and \textbf{gpt-4.1} represent the general-purpose 4.1 series, while \textbf{o4-mini} and \textbf{o3} are reasoning-optimized models with extended thinking capabilities.
All models use default temperature settings with random seeds for each game.
The benchmark automatically captures complete interaction logs, generating comprehensive metrics, efficiency analyses, and visualizations.

\section{Results}
\label{sec:results}

We evaluated five state-of-the-art language models on LemonadeBench v0.5 over 30-day simulations. Our analysis reveals substantial variation in both profitability and operational strategies across models.

\subsection{Overall Performance}

Table~\ref{tab:operational_performance} presents the operational performance metrics across all models. Profit varied dramatically, from \$1,191 (gpt-4.1-nano) to \$13,301 (o3), representing an 11-fold difference in performance on identical tasks.

\input{tables/operational_performance}

\begin{figure}[h]
\centering
\includegraphics[width=0.8\textwidth]{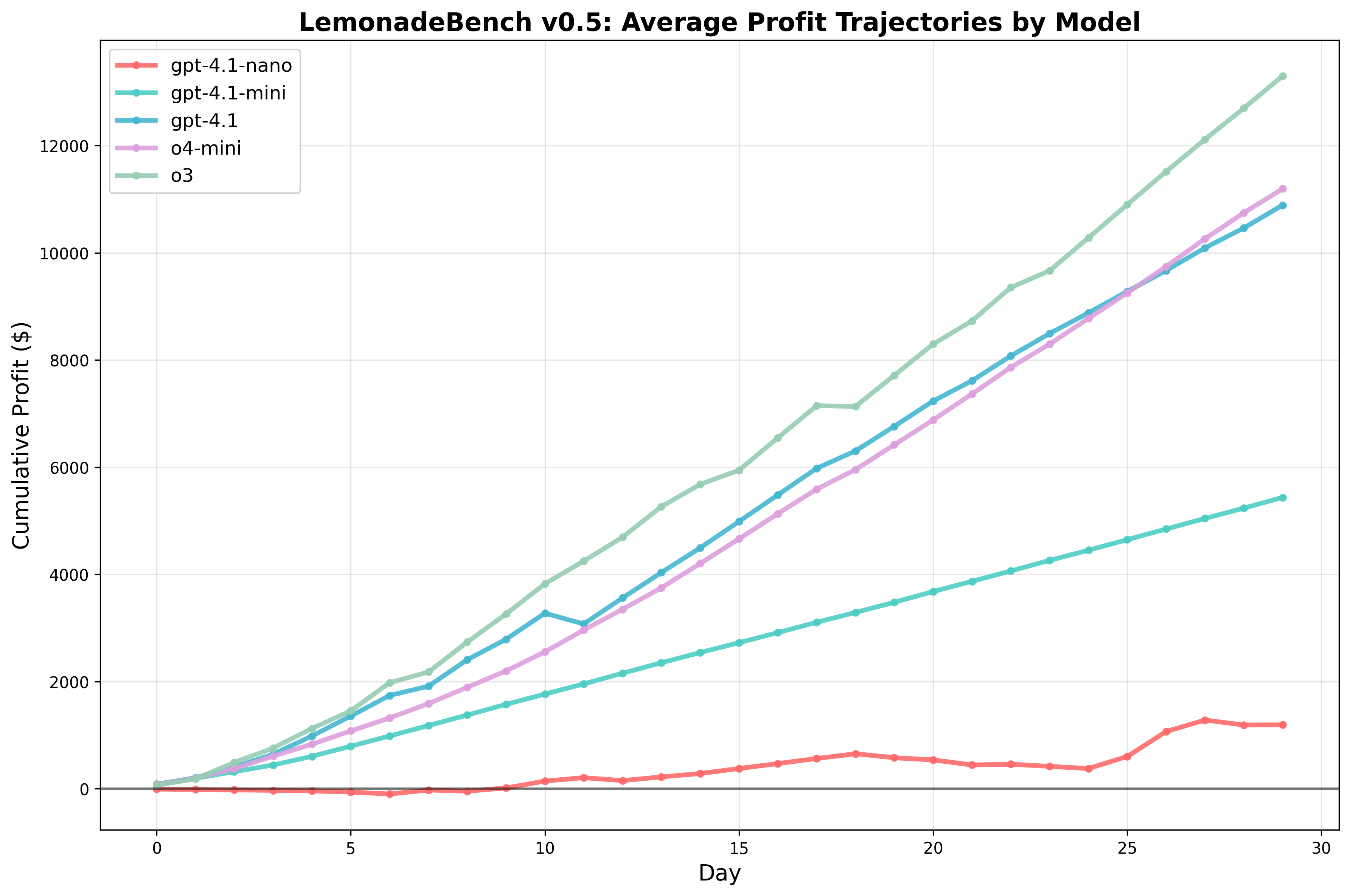}
\caption{Daily profit trajectories showing strategic evolution across models}
\label{fig:profit_trajectories}
\end{figure}

Figure~\ref{fig:profit_trajectories} shows daily profit trajectories across all models, revealing distinct patterns of adaptation and strategic evolution over the 30-day period.

The data reveals distinct strategic approaches: gpt-4.1-mini pursued an ultra-high-volume, low-price strategy (11,450 customers at \$1.00), while o3 implemented a premium pricing strategy (7,941 customers at \$2.24). Notably, o3 achieved the highest profit per customer (\$1.67) and maintained the lowest stockout rate (3/30 days).

\subsection{Business Efficiency Analysis}

Beyond raw profitability, we develop a comprehensive efficiency framework that decomposes performance into six distinct business dimensions. This framework employs counterfactual analysis to isolate the impact of specific decisions, enabling precise diagnosis of where models succeed or fail relative to optimal strategies.

Table~\ref{tab:efficiency_breakdown} presents our efficiency analysis across all models.

\input{tables/efficiency_breakdown}

\subsubsection{Efficiency Metric Definitions}

Our efficiency framework begins with three operational metrics that capture basic execution competence. \textbf{Purchasing Efficiency} measures value captured through strategic timing of supply purchases. With daily costs fluctuating ±10\% around base prices, we calculate:
$$\text{Purchasing Efficiency} = \sum_{i,d} Q_{i,d} \times (C_i^{base} - C_{i,d})$$
where $i$ indexes the four ingredient types (cups, lemons, sugar, water), $d$ represents each day, $Q_{i,d}$ is quantity of item $i$ purchased on day $d$, and $C_{i,d}$ is the actual price paid. Positive values indicate successful market timing—buying when prices dip below base costs.

\textbf{Expired Losses} track the cost of ingredients that spoil before use, particularly critical for lemons which expire after just 7 days:
$$\text{Expired Losses} = -\sum_{i,d} Q_{i,d}^{expired} \times C_i^{base}$$

\textbf{Excess Losses} capture the opportunity cost of ending inventory—capital tied up in unsold goods:
$$\text{Excess Losses} = -\sum_i Q_i^{final} \times C_i^{base}$$

The next three metrics directly evaluate the core business decisions every model must make: how much inventory to maintain, what price to charge, and when to operate. These strategic dimensions reveal the most dramatic performance differences.

\textbf{Stockout Losses} quantify the profit foregone when models turn away customers due to insufficient inventory. This metric asks: if the model had unlimited inventory while maintaining the same prices and operating hours, how much additional profit would it capture? We calculate:
$$\text{Stockout Losses} = -\sum_d C_d^{lost} \times (p_d - c)$$
where $C_d^{lost}$ represents customers turned away on day $d$, $p_d$ is that day's price, and $c$ is the ingredient cost per cup (\$0.37). This counterfactual isolates the pure impact of inventory decisions—revealing that gpt-4.1-nano lost \$9,004 by turning away 80\% of potential customers (5,399 lost sales vs 1,347 served).

\textbf{Pricing Losses} employ counterfactual analysis to isolate revenue impact of pricing decisions. For each day, we ask: if the model maintained its actual inventory levels and operating hours but charged the optimal price (\$2.69), how would profit change? The demand function follows $Q(p) = (50 - 10p) \times \sum_{h \in H_d} m_h$, where $m_h$ is the hourly multiplier. Daily profit is:
$$\pi(p, I, H) = p \times \min(Q(p), I) - c \times \min(Q(p), I) - |H| \times 5$$
where revenue equals price times quantity sold (constrained by demand $Q(p)$ or inventory $I$), minus ingredient costs and operating costs (\$5 per hour). The pricing loss becomes:
$$\text{Pricing Loss}_d = \pi(p_d, I_d, H_d) - \pi(p^*, I_d, H_d)$$
By holding $I_d$ and $H_d$ constant while comparing actual price $p_d$ to optimal price $p^* = \$2.69$, we isolate pure pricing effects. The analysis reveals gpt-4.1-mini's catastrophic \$1.00 pricing strategy cost \$8,127—over 40\% of potential profit.

\textbf{Scheduling Losses} evaluate the opportunity cost of suboptimal operating hours. The globally optimal schedule runs 6am-8pm (hours 6-20), capturing peak demand periods where hourly multipliers range from 0.3 to 1.5. For each day, we calculate total demand under actual vs optimal hours:
$$Q_{actual} = (50 - 10p_d) \times \sum_{h=h_{start}}^{h_{end}} m_h$$
$$Q_{optimal} = (50 - 10p_d) \times \sum_{h=6}^{20} m_h$$
The scheduling loss compares profit under these scenarios:
$$\text{Scheduling Loss}_d = \pi(p_d, I_d, H_d) - \pi(p_d, I_d, H^*)$$
where operating costs change from $(h_{end} - h_{start}) \times 5$ to $15 \times 5 = \$75$ under optimal hours. This reveals whether models effectively exploit their inventory investments—gpt-4.1 lost \$2,422 by operating limited hours despite maintaining adequate stock.

\subsubsection{Model Performance Patterns}

The efficiency breakdown reveals that no model achieves balanced competence across all dimensions.
Instead, each develops a specialized competency profile: gpt-4.1 excelled at purchasing timing (+\$34) but failed at scheduling (-\$2,422).
gpt-4.1-mini captured procurement value (+\$22) but destroyed it through pricing (-\$8,127).
The reasoning models (o3, o4-mini) surprisingly performed worst at market timing (-\$33, -\$37).

Most notably, o3 achieved the highest profit not through excellence in any single dimension but through balanced adequacy across all.
By minimizing the critical stockout problem (\$363 vs gpt-4.1-nano's \$9,004), o3 demonstrated that identifying and addressing primary constraints matters more than optimizing secondary factors.
This strategic focus—enabled by extended reasoning that recognized stockouts as the binding constraint—mirrors how experienced operators allocate scarce attention.

A striking pattern emerges when comparing model performance to theoretical optimal: all models systematically underprice relative to the optimal \$2.69, with average prices ranging from \$1.00 (gpt-4.1-mini) to \$2.24 (o3).
This universal underpricing bias translates directly to performance gaps—models achieve only 6.3\% (gpt-4.1-nano), 28.9\% (gpt-4.1-mini), 57.9\% (gpt-4.1), 59.5\% (o4-mini), and 70.7\% (o3) of the theoretical maximum profit of \$18,811.
The correlation between pricing closer to optimal and overall performance is clear, yet even the best-performing model leaves nearly 30\% of potential profit unrealized through conservative pricing.

\subsubsection{Implications for AI Economic Competence}

These results reveal a fundamental insight: current LLMs develop specialized rather than generalized economic competence.
Each model implicitly prioritizes certain business dimensions while neglecting others, leading to distinctive failure modes.
o3's superior performance suggests that extended reasoning enables better identification of critical constraints.
Rather than pursuing marginal gains in purchasing efficiency or perfect pricing, o3 focused cognitive resources on avoiding the catastrophic stockout problem that crippled other models.

However, our results also reveal a fundamental challenge in benchmark design.
In the current formulation, stockout losses dominate other inefficiencies by an order of magnitude. A model that perfectly times purchases (+\$50) but experiences moderate stockouts (-\$2,000) performs far worse than one with poor purchasing (-\$50) but minimal stockouts (-\$200).
This asymmetry may overweight inventory management relative to real-world importance, potentially rewarding models that simply recognize this imbalance rather than those with genuine economic intuition.

Future versions of LemonadeBench will address these limitations through three approaches: (1) increasing runs per model from 1 to 30 to achieve statistical significance and reduce the impact of randomness in our efficiency measurements, (2) calibrating loss magnitudes using real-world small business data to ensure each dimension reflects actual economic significance, or (3) implementing a balanced scoring system that equalizes the potential impact of each efficiency dimension.
Such refinements will better distinguish between models that possess generalized business acumen versus those that merely exploit the specific reward structure presented.

\subsection{Computational Requirements}

Table~\ref{tab:computational_requirements} shows the computational costs and tool usage patterns for each model.

\input{tables/computational_requirements}

The reasoning models (o3, o4-mini) required significantly more time (32-38 minutes vs 2-8 minutes) and cost (75x more) than standard models, though they achieved superior profits.
o3 costs \$0.97 compared to gpt-4.1-nano's \$0.013, representing a 75x cost increase for an 11x profit improvement.
Both reasoning models made more strategic tool calls (164-184 calls) compared to standard models (90-114 calls), with o3 notably using the \texttt{get\_historical\_supply\_costs()} tool 3 times, indicating deeper market analysis.

\subsection{Reasoning Traces}

The reasoning models produced detailed internal reasoning traces that reveal their strategic evolution over the 30-day period.
We present minimally edited\footnote{Traces are presented verbatim except for formatting improvements: (1) section headers changed from **bold** to \textbf{bold} with em-dash separators for clarity, (2) dollar signs escaped for LaTeX compatibility, (3) Unicode apostrophes replaced with standard ASCII apostrophes, and (4) paragraph breaks added between distinct reasoning segments. No semantic content was altered.} traces from o3's reasoning on days 1, 15, and 30 to illustrate the model's cognitive process:

\textbf{Day 1 - Inventory planning:}
\begin{quote}
\textit{``\textbf{Planning inventory strategy} --- I need to set a strategy for our inventory, starting from likely zero. First, I'll check our current inventory and then the morning prices. My plan is to stock up on shelf-stable items early if the prices are low, but I'm uncertain about how low they might be. It feels like starting from scratch, so I'll keep things flexible as I gather the necessary information. Let's get going!''}

\textit{``\textbf{Checking morning prices} --- I need to check the morning prices first. After that, I can decide about ordering supplies. It's important to optimize these decisions. Right now, I'm not sure about the demand behavior, which makes things a bit trickier. I'll also need to find out what the supply prices are. So, it's essential to look into checking those morning prices to move forward effectively!''}

\textit{``\textbf{Deciding on supply costs} --- I haven't gotten the historical supply costs yet, but I need to determine purchase quantities, prices, and operating hours for Day 1. The goal is to maximize profit over 30 days, starting with \$1000 and a cost of \$5 per hour. Without a demand model, I can't predict how many customers will come based on price or factors like weather. I might set the price based on the assumption of 100 customers per hour at a low price and start by purchasing 200 units of each ingredient.''}

\textit{``\textbf{Calculating lemonade stand setup} --- We've ordered 120 supplies, costing \$44.772, leaving \$955.228 from the starting cash of \$1,000. We need to set our hours and prices before opening the stand. However, we're unsure about customer demand. The baseline ingredient cost per lemonade is \$0.3731, and I've calculated a potential price of \$1.25 or \$1.50. If we operate for 8 hours, our total cost is \$40. At \$1.25, we need to sell 47 cups to break even, which is about 39\% occupancy of our supplies. That seems reasonable!''}
\end{quote}

\textbf{Day 15 - Performance analysis:}
\begin{quote}
\textit{``\textbf{Analyzing sales and profit} --- I have 10 lemons available and can make only 10 cups, so I need to reorder. On day 15, selling lemons at \$2.40 for 10 hours from 9 AM to 7 PM, I sold 275 lemons and made a profit of \$610. The operating cost for those hours is \$50. It's clear that at \$2.40, the demand is strong, but I should consider keeping the price around \$2.50. I'll need a supply of 300 to avoid running out.''}

\textit{``\textbf{Planning for Day 15} --- I'm realizing we need to strategize for Day 15 since our inventory is really low --- only 10 of each ingredient! Historically, we've seen demand around 225-300 sales, peaking at 318. Last time, with a price of \$2.40, we had 275 customers after extending hours. Now, we've got to decide on our pricing strategy. The demand curve shows price variations could drive between 120 and 318 customers, with \$2.40 attracting around 250-275 customers consistently. I need to act quickly to reorder ingredients!''}

\textit{``\textbf{Planning for Day 15 (continued)} --- I'm developing a plan for day 15, where we each have 10 units. To ensure we meet daily inventory needs, we need to examine historical demand; for instance, we sold 275 cups yesterday at a price of \$2.40 over a 10-hour period. Demand appears to decline with price increases. The goal is to maximize profit by balancing revenue, supply, and operating costs. The total variable cost for each lemonade is around \$0.36, while our operating cost for the day amounts to \$50.''}

\textit{``\textbf{Calculating profit insights} --- At a price of \$2.40, earlier customers were 252 (28 per hour). Adding another hour brought in 23 more customers. That's reasonable. Now, I need to compute profits. Revenue equals price multiplied by customers. Operating costs were \$5 per hour for 10 hours. Costs for each lemonade totaled about \$0.3565. When I calculate the profit, it seems they're not accounting for ingredient costs in the recorded profits. It looks like it's already covered in their calculations.''}

\textit{``\textbf{Setting up operating hours} --- I see that the user has ordered supplies and set prices, but they haven't set any operating hours yet. Based on their previous setup, they used 9 AM to 7 PM, which is 10 hours, costing \$50 including profit. That's good! I could consider keeping the same hours or possibly adjusting them, but I need to be mindful that lemons are the limiting factor, allowing for only 350 cups. So, I need to choose hours expecting 350 customers or less.''}
\end{quote}

\textbf{Day 30 - Final optimization:}
\begin{quote}
\textit{``\textbf{Analyzing lemonade inventory and sales} --- Today is the final day, and I have an inventory of 591 cups, 341 lemons, 592 sugars, and 592 waters. Since lemons are the limiting factor, I can make 341 lemonades. Pricing and hours need to be decided; historically, I charged \$2.40 and operated 10 hours from 9 AM to 7 PM. Given that I had around 263 customers yesterday, I could expect similar demand today. Maximizing profit is key, even with leftover inventory.''}

\textit{``\textbf{Considering final day operations} --- I see that we have 341 cups available and the price is set at \$2.40. With historical demand around 270 customers for Day 30, we should be fine. Operating hours are currently from 9 AM to 7 PM, but I'm pondering whether to extend by starting an hour earlier. This could potentially attract 10 to 15 more customers, but there's an added cost of \$5 per hour. It's all about weighing the profit against that extra operating cost, isn't it?''}
\end{quote}

The reasoning traces reveal a fascinating progression in o3's economic understanding over the 30-day period. On Day 1, the model operates under extreme uncertainty, making conservative assumptions (100 customers per hour at low prices) and focusing on basic break-even calculations. By Day 15, o3 has developed a sophisticated understanding of demand elasticity, recognizing that \$2.40 pricing consistently attracts 250-275 customers and explicitly considering price optimization around \$2.50. The model now thinks in terms of demand curves and marginal analysis, calculating per-unit costs (\$0.3565) and hourly customer rates (28 per hour).

By Day 30, o3 demonstrates mature economic reasoning with explicit cost-benefit analysis of operating hour extensions. The model's consideration of marginal revenue (``10 to 15 more customers'') versus marginal cost (``\$5 per hour'') shows genuine understanding of optimization at the margin. This evolution from naive quantity assumptions to sophisticated price discovery and finally to marginal analysis mirrors how human entrepreneurs learn through experience, suggesting that extended reasoning enables emergent economic intuition.

\section{Conclusion}
\label{sec:conclusion}

LemonadeBench v0.5 reveals both the promise and limitations of current AI systems in economic decision-making. While all models achieve profitability—with the best capturing 70\% of theoretical optimal—they do so through divergent strategies that expose fundamental gaps in economic reasoning. The universal tendency to underprice, combined with specialized competencies that excel in some dimensions while failing in others, suggests that current models learn locally optimal heuristics rather than globally optimal strategies.

\subsection{Reflections on AI-Assisted Research}

This paper itself serves as a case study in human-AI collaboration. We extensively used Claude Code, OpenAI's o1 models, Claude Opus 4, Claude Sonnet 4, and o3 throughout the research process. These tools undeniably accelerated development—debugging efficiency calculations that would have taken hours required only minutes.

Yet the limitations were equally instructive. The AI assistants struggled with both conceptualizing and implementing economic logic. Not one suggested the counterfactual efficiency framework that became our key analytical contribution. They required extensive handholding and constant guidance to maintain focus on the research questions. The models were unable to generate any of the main ideas of this paper, despite having access to all the data and context. While these tools were certainly helpful and sped up coding and writing time, they remained fundamentally limited to executing well-defined tasks within a human-provided framework.

\subsection{Research Roadmap}

This initial benchmark represents only the beginning of a comprehensive research program. Computational constraints limited v0.5 to a simplified 30-day simulation, but our vision extends far beyond:

\textbf{LemonadeBench 1.0} will expand to decade-long simulations with 30 runs per model for statistical significance. This version will introduce realistic business complexity: decisions between automation and human labor, marketing initiatives with uncertain ROI, multi-location expansion strategies, capital structure optimization through debt issuance and stock buybacks, vertical integration opportunities, and seasonal demand patterns reflecting holidays and weather. These additions transform the benchmark from a simple optimization problem to a genuine test of strategic business thinking.

\textbf{LemonadeBench 2.0} will shift from single-agent to multi-agent economics, implementing a global demand function where multiple AI-operated lemonade stands compete directly. We envision three experimental conditions: (1) a baseline where models operate independently, (2) a communication-enabled version to observe whether models spontaneously form cartels or engage in tacit collusion, and (3) a legally-constrained version where models must balance profit maximization with explicit prohibitions against anti-competitive behavior. This final condition directly tests a critical concern for AI deployment: can models navigate the tension between economic optimization and legal compliance?

Throughout these expansions, we will incorporate human baseline performance to ground our evaluations in realistic expectations. We actively seek partnerships with major AI labs to develop this comprehensive benchmark suite, recognizing that the computational and engineering requirements exceed what any single researcher can accomplish.

\section*{Code and Data Availability}

The LemonadeBench benchmark, evaluation code, and experimental results are available at \url{https://github.com/aidanvyas/lemonadebench}.

\section*{Acknowledgments}

We thank OpenAI for providing complimentary tokens through their data sharing program, which enabled comprehensive evaluation across multiple models at a reduced cost.

\bibliographystyle{unsrtnat}
\bibliography{references}

\end{document}

%% file: tables/operational_performance.tex
\begin{table}[h]
\centering
\caption{Operational Performance Metrics}
\label{tab:operational_performance}
\resizebox{\textwidth}{!}{%
\begin{tabular}{|l|r|r|r|r|r|r|r|}
\hline
\textbf{Model} & \textbf{Profit (\$)} & \textbf{Customers} & \textbf{Avg Price (\$)} & \textbf{Revenue (\$)} & \textbf{Hours Open} & \textbf{Stockout Days} & \textbf{\$/Customer} \\
\hline
gpt-4.1-nano & 1,191 & 1,347 & 2.22 & 2,996 & 240 & 29/30 & 0.88 \\
\hline
gpt-4.1-mini & 5,435 & 11,450 & 1.00 & 11,450 & 360 & 30/30 & 0.47 \\
\hline
gpt-4.1 & 10,890 & 7,155 & 2.06 & 14,748 & 238 & 6/30 & 1.52 \\
\hline
o4-mini & 11,195 & 9,085 & 1.78 & 16,131 & 300 & 7/30 & 1.23 \\
\hline
o3 & 13,301 & 7,941 & 2.24 & 17,765 & 285 & 3/30 & 1.67 \\
\hline
\end{tabular}
}%
\end{table}

%% file: tables/efficiency_breakdown.tex
\begin{table}[h]
\centering
\caption{Business Efficiency Breakdown}
\label{tab:efficiency_breakdown}
\resizebox{\textwidth}{!}{%
\begin{tabular}{|l|r|r|r|r|r|r|r|}
\hline
\textbf{Model} & \textbf{Profit (\$)} & \textbf{Purchasing (\$)} & \textbf{Expired (\$)} & \textbf{Excess (\$)} & \textbf{Stockout (\$)} & \textbf{Pricing (\$)} & \textbf{Scheduling (\$)} \\
\hline
gpt-4.1-nano & 1,191 & -3 & 0 & -103 & -9,004 & -263 & -97 \\
\hline
gpt-4.1-mini & 5,435 & +22 & 0 & 0 & -1,761 & -8,127 & -277 \\
\hline
gpt-4.1 & 10,890 & +34 & -55 & 0 & -619 & -871 & -2,422 \\
\hline
o4-mini & 11,195 & -37 & 0 & -37 & -481 & -2,737 & -1,445 \\
\hline
o3 & 13,301 & -33 & 0 & -68 & -363 & -647 & -1,057 \\
\hline
\end{tabular}
}%
\end{table}

%% file: tables/computational_requirements.tex
\begin{table}[h]
\centering
\caption{LemonadeBench v0.5 - Computational Requirements and Tool Usage}
\label{tab:computational_requirements}
\resizebox{\textwidth}{!}{%
\begin{tabular}{|l|r|r|r|r|r|r|r|r|r|r|r|}
\hline
\textbf{Model} & \textbf{Profit (\$)} & \textbf{Total} & \textbf{Check} & \textbf{Check} & \textbf{Order} & \textbf{Set} & \textbf{Set} & \textbf{Historical} & \textbf{Open} & \textbf{Time (s)} & \textbf{API Cost (\$)} \\
 & & \textbf{Calls} & \textbf{Inv.} & \textbf{Prices} & \textbf{Supplies} & \textbf{Price} & \textbf{Hours} & \textbf{Data} & \textbf{Business} & & \\
\hline
gpt-4.1 & 10,889 & 172 & 30 & 30 & 22 & 30 & 30 & 0 & 30 & 493.7 & 0.3080 \\
\hline
gpt-4.1-mini & 5,435 & 180 & 30 & 30 & 30 & 30 & 30 & 0 & 30 & 189.6 & 0.0637 \\
\hline
gpt-4.1-nano & 1,191 & 180 & 30 & 30 & 30 & 30 & 30 & 0 & 30 & 114.6 & 0.0130 \\
\hline
o3 & 13,300 & 164 & 30 & 25 & 16 & 30 & 30 & 3 & 30 & 1902.7 & 0.9710 \\
\hline
o4-mini & 11,195 & 184 & 30 & 30 & 32 & 30 & 30 & 2 & 30 & 2267.4 & 0.8804 \\
\hline
\end{tabular}
}%
\end{table}